Development of competitive high-entropy alloys using commodity powders

José M. Torralba (1,2,*), S. Venkatesh Kumarán (2)

(1) Universidad Carlos III de Madrid, IMDEA Materials Institute, 28911 Leganes, Madrid, Spain
(2) IMDEA Materials Institute, 28906 Getafe, Madrid, Spain

*Corresponding author

**Abstract**

One of the main drawbacks of the powder metallurgy route for High-Entropy Alloys (HEAs) is the unavailability of fully pre-alloyed powders in the market. Using commodity powders (commercial Ni, Fe and Co base fully pre-alloyed powders, fully available in large quantities and at competitive prices) to produce HEAs presents a completely new and competitive scenario for obtaining viable alloys for high-performance applications.

1. **Introduction.**

Since the concept of "high-entropy alloys" (HEAs) was introduced in 2004 [1,2], extensive research on this family of alloys has resulted in more than 4000 papers, according to the Scopus database. Despite ingot metallurgy is the most common way to develop and manufacture HEAs, powder metallurgy (PM) has shown high potential for manufacturing HEAs. PM has provided an alternative technology for HEA development that achieves significantly higher compositional accuracy than other methods, while preventing segregation and achieving superior microstructural control [3]. Powder metallurgy high-entropy alloys (PMHEAs) have been developed using three different classes of powders to date: fully preallyoed powders (usually gas-atomized), pure elemental powders and mechanically alloyed powders (which are fabricated from elemental powders) The source of the powders depends on the PM manufacturing route selected. The cost and availability of powders limit the use of PM for fabricating HEAs. There are no fully preallyoed powders on the market with the specific compositions of even the most extensively studied HEAs and pure powders are usually expensive and can be difficult to manage. However, many grades of powder are available on the market that belong to the families of metals on which many HEAs are based: Ni, Cr, Fe, Co, Ti, Al, etc. These grades are all fully commercially available (the compositions of these powders are widely used in the industry) from several manufacturers and can be delivered in large quantities within short times at competitive prices. These families of available alloys in mass production are designated by the term "commodity", for which the following hypothesis is presented in this paper: is it feasible to produce a HEA using a mix of commodity powders as raw materials? We investigate this hypothesis by predicting the HEA formation from mixing properly fully preallyoed commodity powders Ni625, INVAR36, CoCrF75 (62%Co-32%Cr-4%Mo-Co bal.wt.%), Fe49Ni and 316L. We also validate the hypothesis by manufacturing by PM route one of the proposed compositions.

2. **Materials and methods.**

The selected alloys are available on the market as gas-atomized powders from different providers (the Co base powders from VDM Metals (Germany) and the rest from Sandvick Osprey (UK). Table 1 shows the proposed weight percentages mixed in the alloys to produce the final composition, shown in atomic percentages. Can a "real" HEA be obtained using this powder mixes?

This novel class of materials known as high-entropy alloys were first defined as those containing five or more elements in relatively high (5–35 at.%) or even equiatomic concentrations [1,4]. This composition-based definition was later modified [5]. That is, HEAs are no longer required to be equimolar and can contain minor elements used to modify the properties of the base HEA [6]. Our proposed HEA based on three commodity powders would not fall within existing definitions of HEAs. However, a definition based on the magnitude of the entropy has been considered since the earliest development of these alloys. The configurational mixing entropy $\Delta S_{mix}$ can be used to classify alloys as low ($\Delta S_{mix, ideal} < 0.69R$), where $\Delta S_{mix, ideal}$ is the total configurational molar entropy in an ideal solid solution and R is the gas constant), medium ($0.69R < \Delta S_{mix, ideal} < 1.61R$) and high-($\Delta S_{mix, ideal} > 1.61R$) entropy alloys[6]. The value of the configurational mixing entropy ΔSmix can be easily obtained [7]. Also the value of the enthalpy of mixing for a multicomponent alloy system with n elements can be determined to predict the formation of the HEA [8].

Table 1. Proposed mixes of commodity powders (and role of each powder) used to develop different possible HEAs.

|  | Weight % | | | | | Atomic % | | | | |
|---|---|---|---|---|---|---|---|---|---|---|
| Alloy | Ni625 | INVAR 36 | CoCrF75 | 316L | Fe49Ni | Ni | Fe | Cr | Mo | Co |
| C1 | 20 | 38 | 42 | | | 25,45 | 27,42 | 20,44 | 1,59 | 25,1 |
| C2 | 28 | | 38 | 33 | | 21.18 | 24.42 | 25.92 | 3.6 | 23.11 |
| C3 | | | 48 | | 52 | 24.44 | 27.55 | 15.63 | 1.79 | 29.96 |
| Role (source of) | Ni, Cr, Fe and Mo | Fe and Ni | Co, Cr and Mo | Fe, Cr, Ni and Mo | Fe, Ni | | | | | |

Other criteria have been introduced to predict the formation of a single solid solution (SSS) HEA [9]. A recent study provided an extensive classification of such alloys, including an analysis of the compositional accuracy and the factors that significantly impact this accuracy[10], including the parameters, Ω and δ, defined in [18] to assess the forming ability of a simple solid solution. Ω reflects the relative strength of the entropy and the enthalpy, and δ is a measure of the atomic size difference in the alloy. Thus, a simple solid solution forms when the entropy is large relative to the enthalpy and there is a small atomic size difference in the alloy. A novel parameter, γ, was proposed in [11] as a measure of the atomic size difference[12]. Singh et al.[13] proposed another single-parameter model by defining $\Lambda = \Delta S_{mix}/\delta^2$. SSS formation was reported to be favored for Λ> 0.96, i.e., a large configurational entropy and a small atomic size difference. Atomic size differences play a central role in this model.

The valence electron concentration (VEC) can be used to determine the suitability of an FCC or BCC alloy as an HEA [14]. The VEC plays a decisive role in determining whether an HEA exists as a FCC- or BCC-type solid solution, that is, a large VEC (≥8) favors the formation of FCC-type solid solutions, whereas a small VEC (<6.87) favors the formation of BCC-type solid solutions.

Once we assess the prediction that the proposed alloy compositions can form a simple solid solution, we have fabricated one of the alloys (C1) as a bulk material by manual mixing followed by field-assisted hot pressing and annealing. The three commodity powders were used as the raw materials (the quantity of each powder in the mix are shown in Table 1). Field-assisted hot pressing was performed using a Gleeble3800 at 50 MPa and 1000 °C for 15 minutes. The obtained density of the sintered samples was 94% of the theoretical density. Annealing was performed in vacuum at 1200 °C for 24 hours. A phase analysis was performed using the PANalytical X-ray diffractometer with a Cu Kα source. The microstructures were observed via double-beam scanning electron microscopy (Helios Nanolab 600i) equipped with an electron backscatter diffraction (EBSD) device.

3. Results and discussion.

Table 2 shows the calculated values of the aforementioned parameters for the proposed alloys, as well as the threshold values. The calculated parameter values for the proposed alloys suggest that PMHEAs can be feasibly obtained using mixes of commodity powders, and according these predictions the alloys C1 and C3 will produce a FCC single phase and C2 a mix of BCC and FCC.

Figure 1 (down) shows an EBSD inverse pole figure (IPF) map of the as-sintered alloy. The average grain size was measured as 1.72µm. The EBSD phase map shows three main phases (the FCC phase is shown in red, the HCP phase is shown in green and the BCC phase is shown in blue), which were also detected by XRD (Figure 1, up).

Table 2. Different assessment parameters for HEAs.

| Parameter | Obtained value | | | Threshold values | Reference |
|---|---|---|---|---|---|
| | C1 | C2 | C3 | | |
| $\Delta S_{mix}^{conf}$ | 1.44R | 1.48R | 1.43R | > 1.61R | [6] |
| $\Delta H_{mix}$ (kJ/Mol) | -6.3727 | -6.2822 | -6.636 | -11.6<$\Delta H_{mix}$<3.2 | [15] |
| $\Omega/1000$ | 3.52 | 3.7 | 3.3 | ≥1.1 | [16] |
| $100\delta$ | 1.69 | 2.24 | 1.85 | <6.6 | [16] |
| γ | 1.1324 | 1.1333 | 1.1328 | <1,175 | [11] |
| Λ | 4.19 | 9.58 | 9.21 | >0.96 | [13] |
| VEC | 8.32 | 7.92 | 8.38 | >8 for FCC | [17] |

The HEA was fully developed after annealing (Figure 2), and only the FCC phase can be observed in both the XRD pattern and the EBSD phase map. The microstructure shows only FCC grains (Figure 2, down). The average grain size in the annealed sample is 12.26 µm. There is an slightly increase in the grain size due to the higher annealing temperature/time.

4. **Conclusions.**

The results of this preliminary study show that powder metallurgy presents a promising way to produce HEAs using commodity powders, which are inexpensive raw materials (in comparison with gas-atomized powders tailored to specific HEA compositions or elemental powders) that are fully commercially available in large quantities. This study identifies new opportunities for the development of HEAs and their implementation in industry.

**Aknowledgments**

Authors wish to thank to IMDEA Materials Institute for the financial support to this research.

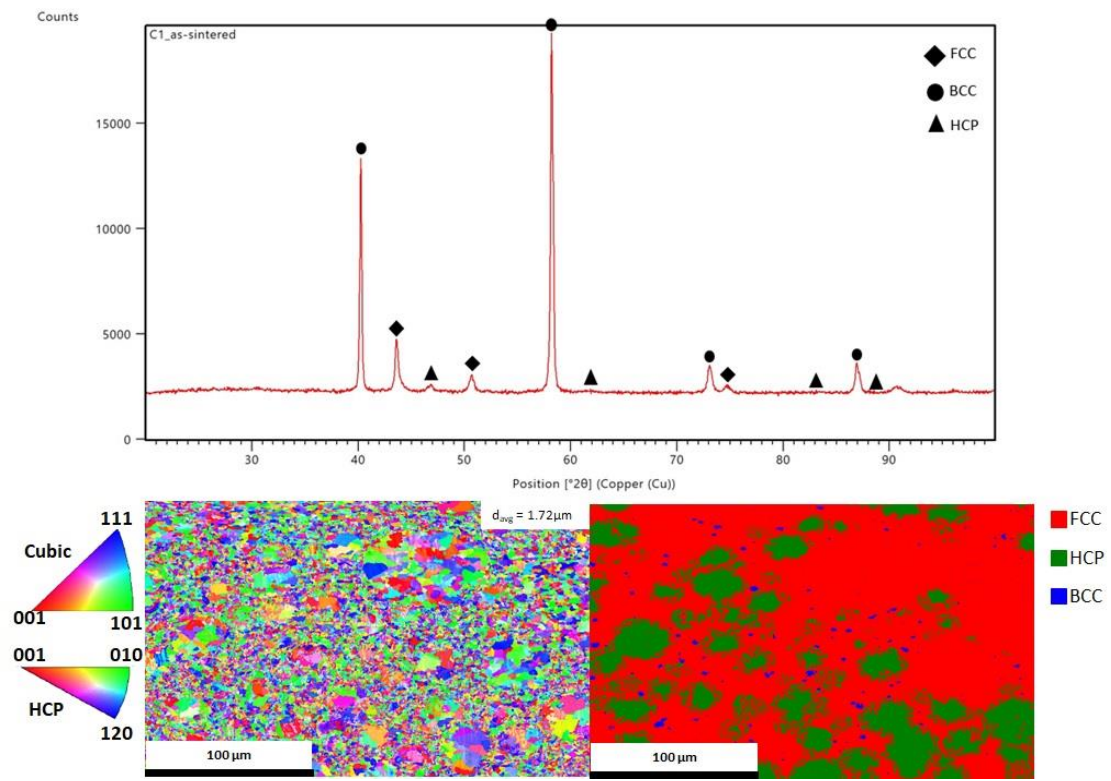

Figure 1. Up, XRD pattern and, down, EBSD inverse pole figure (IPF-Z) map of the alloy and EBSD phase map of the as-sintered sample.

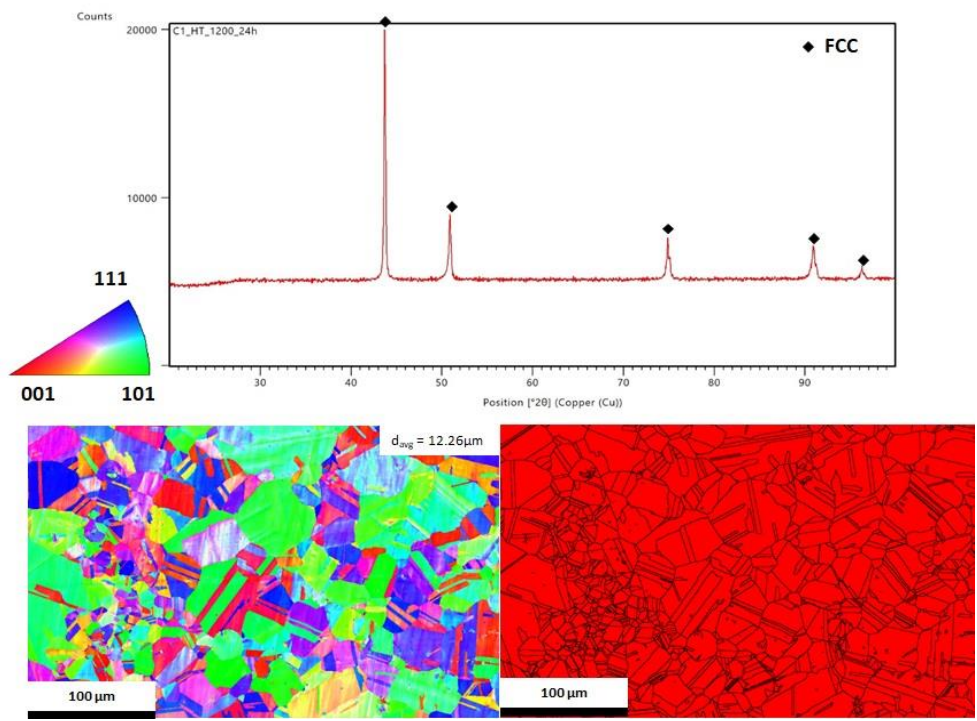

Figure 2. Up, XRD pattern of the as-sintered sample. Down, EBSD inverse pole figure (IPF-Z) map of the alloy and EBSD phase map.